\documentclass[letterpaper,twoside,12pt]{article}
\usepackage{amssymb}
\usepackage{amsmath}
\usepackage{latexsym}
\usepackage{verbatim}
\usepackage{graphicx}
\usepackage{epsfig}
\usepackage{stmaryrd}
\usepackage{rotating,graphicx}
\usepackage[english]{babel}
\usepackage{setspace}
\usepackage[english]{babel}
\usepackage[utf8]{inputenc}
\usepackage[T1]{fontenc}
\usepackage{lmodern}

\begin{document}

\newcommand{\bra}[1]{\langle #1|}
\newcommand{\ket}[1]{|#1\rangle}
\newcommand{\braket}[2]{\langle #1|#2\rangle}

\begin{Large}
\begin{center}
\textbf{On the Consistency of the Consistent Histories Approach to Quantum Mechanics}\\
\end{center}
\end{Large}

\begin{center}
\begin{large}
Elias Okon\\
\end{large}
\textit{Instituto de Investigaciones Filosóficas, Universidad Nacional Autónoma de México, México City, México.\\ E-mail: eokon@filosoficas.unam.mx}\\[.5cm]
\begin{large}
Daniel Sudarsky\\
\end{large}
\textit{Instituto de Ciencias Nucleares, Universidad Nacional Autónoma de México, México City, México.\\ E-mail: sudarsky@nucleares.unam.mx}\\[.5cm]

\end{center}
\noindent \textbf{Abstract:}
The Consistent Histories (CH) formalism aims at a quantum mechanical framework which could be applied even to the universe as a whole. CH stresses the importance of histories for quantum mechanics, as opposed to measurements, and maintains that a satisfactory formulation of quantum mechanics allows one to assign probabilities to alternative histories of a  quantum system. It further proposes that each realm, that is, each set of histories to which probabilities can be assigned, provides a valid quantum-mechanical account, but that different realms can be mutually incompatible. Finally, some of its proponents offer an ``evolutionary'' explanation of our existence in the universe and of our preference for quasiclassical descriptions of nature. The present work questions the validity of claims offered by CH proponents asserting that it solves many interpretational problems in quantum mechanics. In particular, we point out that the interpretation of the framework leaves vague two crucial points, namely, whether realms are fixed or chosen and the link between measurements and histories. Our claim is that by doing so, CH overlooks the main interpretational problems of quantum mechanics. Furthermore, we challenge the evolutionary explanation offered and we critically examine the proposed notion of a realm-dependent reality.

\onehalfspacing
\section{Introduction}
The Consistent Histories approach to quantum mechanics, developed by Griffiths \cite{Gri:84,Gri:86,Gri:87}, and further elaborated by Gell-Mann and Hartle \cite{Gel.Har:90,Har:91,Gel.Har:93}, and Omnès \cite{Omn:87,Omn:88}, aims at a quantum-mechanical framework which, in sharp contrast with the standard interpretation of quantum mechanics, dispenses with the notion of measurement and the distinction between the observer and what is observed; (contemporary presentations of the formalism include \cite{Gri:03,Gri:09,Hoh:10,Gri:11,Gri:13}). In particular, it is a proposal for a formulation of quantum mechanics applicable to the universe as a whole. CH stresses the importance of histories for quantum mechanics, as opposed to measurements, and posits that a satisfactory formulation of quantum mechanics should be one that allows the assignment of probabilities to alternative histories of a  quantum system. The central problem that it must overcome, however, is that not all histories can be assigned probabilities so it must provide an observer-independent criterion for deciding which sets of histories can be so endowed, supplying, for the appropriate cases, rules to compute the corresponding probabilities.

The subject of this paper will be the particular version of CH introduced by Gell-Mann and Hartle in \cite{Gel.Har:90,Har:91,Gel.Har:93} and further developed and applied in \cite{Gel.Har:94,Har:94,Har:96,Har:98,Gel.Har:98,Gri.Har:98,Har:98a,Har:05,Har:07,Har:07a,Gel.Har:07,Har:08,Har:11,%
Gel.Har:11,Har:13}; following \cite{Gri:13}, we will refer to such version of CH as the Decoherent Histories Interpretation (DHI).\footnote{It is important to note that DHI has not suffered significant changes along these 23 years, as can be appreciated from comparing its description in, e.g., \cite{Gel.Har:90} and \cite{Har:11}.} It is our opinion that, in spite of profound similarities, DHI is significantly different in key respect from other approaches to CH (e.g., Griffiths') in order to deserve an individualized evaluation. In fact, it differs most precisely in those conceptual and interpretational issues we are interested in analysing in this work: the role of information gathering and utilizing systems (IGUSes)  in the formalism, positions regarding the nature of reality and an emphasis on applications to cosmology; (a detailed critique of Griffiths' version of the formalism, in particular with respect to its treatment of measurements, is presented in the companion paper \cite{Oko.Sud:13}). Moreover, we find it relevant to evaluate this particular interpretation of the formalism because: i) it is one of the  approaches that continues to be taken seriously by a sizeable sector of the community, and, more importantly, ii) despite the fact that, over the last 23 years, many criticism of this version of CH have been raised, to the best of our knowledge the objections presented in these paper are novel.

DHI relies heavily on the notion of a \emph{realm}, i.e., a set of coarse-grained histories to which probabilities can be assigned. According to DHI, each realm provides a valid quantum-mechanical account of the historical development of a given system. However, the peculiarities of quantum theory often allow for different realms to be mutually incompatible. The view is then that quantum-mechanical statements are meaningful only relative to a particular realm. Moreover, the approach maintains that the standard, Copenhagen-type, measurement situation is nothing more than a special case of a setting for which the CH formalism allows for probabilities to be assigned. Consequently, a central claim of the proposal is that the standard interpretation of quantum mechanics must be seen as a limiting case of the CH framework. DHI also uses the CH formalism to offer an ``evolutionary'' explanation for the existence in the universe of complex adaptive systems, and, in particular, of IGUSes, like humans, and for their almost exclusive preference for quasiclassical descriptions of nature.

The present work offers a critical assessment of these ideas and in particular of claims explicitly made in, for example, \cite{Gel.Har:90,Har:91,Har:93}, asserting that DHI accomplishes the resolution of many of the problems of interpretation present in standard quantum mechanics. Specifically, we point out that the interpretation proposed in these works leaves unresolved two crucial points, namely, whether humans and other IGUSes can ``choose'' or not a realm, and the link between measurements and histories. As such, the proposal ends up overlooking the main interpretational problems of quantum mechanics. Moreover, we will argue that there seems to be no possible satisfactory resolution of those issues within CH. Furthermore, we will critically examine the evolutionary explanation offered and the proposal for a realm-dependent reality.

The plan for this manuscript is as follows. Section \ref{CH} presents the CH formalism and section \ref{GHI} describes its interpretation as developed in  DHI. Section \ref{MO} then reviews the main objections that have been raised against CH and section \ref{RO} presents four new objections that we take to signal severe remaining conceptual problems in the DHI formulation of quantum mechanics. Section \ref{Conc} closes with some final thoughts.

\section{Consistent Histories}
\label{CH}
The notion of measurement and the distinction between the observer and what is observed play central roles in the standard interpretation of quantum mechanics. These features render such interpretation unfit to be applied to closed systems (e.g. ones that include observers) and, in particular, inadequate for clear applicability of quantum theory to cosmology. The CH formulation of quantum mechanics, in contrast, aims at a quantum-mechanical framework for closed systems and, specifically, at one that is applicable to the universe as a whole. The formalism is also considered by its proponents as a completion of the Everettian program, or as the ``right way'' to develop Everett's ideas. 

The CH formalism holds that the most general objective of quantum mechanics is the prediction of probabilities for time-histories of a system. In order to achieve this, it provides an observer-independent criterion to tell what sets of alternative histories of a given system can be assigned probabilities and allows for these probabilities to be computed. Lets see how all these is done.

The formalism takes as inputs an initial state $ \ket{\psi} $ and some dynamics dictated by a Hamiltonian operator $ \hat{H} $. These are supposed to be given by an external theory and, in the case of cosmology, by a fundamental cosmological theory (presumably consistent with a full quantum theory of gravity). The formalism then introduces the notion of an exhaustive and exclusive set of yes/no alternatives (or facts) at a time. Such sets are represented, in the Heisenberg picture, by a set of projection operators:
\begin{equation}
\lbrace P_\alpha (t) \rbrace , \qquad \alpha =1,2,3,...
\end{equation}
such that 
\begin{equation}
{\sum}_\alpha P_\alpha (t) =1, \qquad \mbox{and} \qquad P_\alpha (t) P_\beta (t) = \delta_{\alpha\beta} P_\alpha (t) .
\end{equation}
The first of the equations above implements the exhaustiveness of these projections, the second the exclusiveness. From these sets, the notion of a \emph{set of histories}, which is a time-sequence of sets of (exhaustive and exclusive) facts, is constructed. The sets of histories are represented by 
\begin{equation}
\lbrace P^1_{\alpha_1} (t_1) \rbrace, \lbrace P^2_{\alpha_2} (t_2) \rbrace,..., \lbrace P^n_{\alpha_n} (t_n) \rbrace , \quad \mbox{at times} \quad t_1<t_2<...<t_n .
\end{equation}
Individual histories are then assembled by selecting a particular sequence of alternatives, $(\bar{\alpha}_1,\bar{\alpha}_2,...,\bar{\alpha}_n)$, one of each set. Such histories are represented by the corresponding chain of projections $C_{\bar{\alpha}} \equiv P^n_{{\bar{\alpha}}_n} (t_n)...P^1_{{\bar{\alpha}}_1} (t_1)$ and each history gets assigned a branch state vector: $ \ket{\psi_{\bar{\alpha}}} = C_{\bar{\alpha}} \ket{\psi} $.

Sets of histories are generally coarse-grained because alternatives are not specified at all times and because the projections involved can be projections onto subspaces of dimension greater than one. Operations of fine- and coarse-graining can be defined on sets of histories by, for example, removing or adding sets of alternatives, or by combining or refining the projections.

The next step in the formalism is to assign probabilities to individual histories within a set, but it turns out that not all sets of histories can be assigned probabilities. That can be done only when there is \emph{negligible} interference between branches of the set
\begin{equation}
\langle \psi_\alpha| \psi_\beta\rangle \approx 0,
\end{equation}
a condition that ensures that the assigned probabilities (approximately) satisfy the axioms of probability. Sets satisfying the condition above are said to (medium) {\it decohere}. According to the formalism, these are the only sets for which quantum mechanics makes predictions, with (approximate) probabilities for different branches given by $ p(\bar{\alpha}) = \Vert C_{\bar{\alpha}} \ket{\psi} \Vert^2 $. A decoherent set of alternative coarse-grained histories is known as a realm.

With the formalism in place, one can now extract information from the theory. Then, given data $d$ at time $t_0$, represented by a projection operator $P_d (t_0)$, along with the initial state $ \ket{\psi} $ and the Hamiltonian $ \hat{H} $, predictions for the probability of the future history $\alpha_f$ are given by 
\begin{equation}
\label{pre}
p(\alpha_{f} | d) = \dfrac{\Vert C_{\alpha_{f}} P_d (t_0)\ket{\psi} \Vert^2} {\Vert P_d (t_0)\ket{\psi} \Vert^2} ,
\end{equation}
with $C_{\alpha_{f}}$ an exhaustive set of alternative histories to the future of $t_0$. Similarly, retrodictions for the past history $\beta_p$ are given by 
\begin{equation}
\label{ret}
p(\alpha_{f} | d) = \dfrac{\Vert P_d (t_0) C_{\beta_{p}} \ket{\psi} \Vert^2} {\Vert P_d (t_0)\ket{\psi} \Vert^2} ,
\end{equation}
with $C_{\beta_{p}}$ an exhaustive set of alternative histories to the past of $t_0$.

Recapitulating, the most important features of the CH formalism are i) the fact that it uses histories, as opposed to instantaneous states, as central descriptive tools for the theory; ii) that it implements temporal evolution \emph{only} via Schr\"odinger's dynamics, without (at least explicit) mention of the projection (or collapse) postulate; and iii) that it provides an observer-independent criterion for deciding which sets of histories can be assigned probabilities and gives rules to tell what these probabilities are.
\section{An Interpretation of the Formalism}
\label{GHI}
In this section we will explore four core aspects of DHI: the notion of incompatible realms, the relation between the concept of a quasiclassical realm and the existence of complex creatures such as ourselves, the way in which the formalism is supposed to imply the standard interpretation of quantum mechanics and the proposal for a relativism of reality. We will discuss these in order.

\subsection{Inconsistent realms}
As we saw in the previous section, a realm is a set of histories for which probabilities can be consistently assigned. It turns out that, given a generic system, many different realms can be defined. Furthermore, the theory does not distinguish between all these different realms; it treats all of them on an equal footing. However, not all realms are compatible in the sense that two different realms of the same system may lead to contradictory conclusions.

Lets see how this works in detail. We start by defining two realms as \emph{incompatible} if there is no common finer-grained realm (which by definition must exclude non-negligible interferences) of which they are both coarse-grainings. Then, it can be shown (see \cite{Ken:97}) that using two incompatible realms, both compatible with the same given data, it is possible to arrive at inconsistent stories of what actually happened. That is, it is possible to retrodict, with \emph{certainty} in each realm, two inconsistent facts about the past.
 Therefore, one is forced to conclude that, according to CH, there is no unique past given present data.

DHI clearly recognize this complication, and in order to avoid inconsistencies impose the following rule: \emph{inferences may not be drawn by combining probabilities from incompatible realms} (see \cite{Gri.Har:98}). Making such kind of deductions is just something you are not allowed to do while using the formalism; (Griffiths calls this rule the Single-Framework Rule, \cite{Gri:03}).
\subsection{Quasiclassical domains and IGUSes}
\label{QI}
A \emph{quasiclassical domain} is defined in \cite{Gel.Har:90} as a realm that is maximally refined (in the sense that if you further fine-grain it, it ceases to decohere) and that contains individual histories exhibiting as much patterns of classical correlation in time as possible (see also \cite{Har:94,Har:07a,Har:11}). The world we perceive is supposed to be the foremost example of such a domain. In addition, humans are taken to be complex adaptive systems, and, in particular, special types of IGUSes (see \cite{Gel.Har:90,Gel.Har:94,Har:98,Har:98a,Har:07}). The most important characteristic of an IGUS is considered to be the fact that it uses a (maybe rudimentary) physical theory in order to make predictions about its surrounding environment.

With these two ideas in place, it is sustain that the existence of IGUSes, and  more importantly their  reliance on the semiclassical realms, is to be explained in \emph{evolutionary} terms: they evolved to make predictions because it is adaptive to do so and they focus on quasiclassical domains because these present enough regularity to permit predictions by rudimentary methods. Then, this is supposed to explain why, among all the possible realms that the CH formalism allows, we as humans experience only a very particular type, namely a quasiclassical one.\footnote{It is not clear if there exits just one quasiclassical realm. If more than one exists we should ask whether different IGUS of classes of IGUSes could possibly perceive different ones.}
\subsection{Recovering the standard interpretation}
In order to show that the standard interpretation of quantum mechanics is contained in the CH formulation, reference \cite{Gel.Har:90} starts by defining a \emph{measurement} as a correlation between values of operators of a quasiclassical domain. Then, the claim is that this implies that the standard, Copenhagen-type, measurement situation, i.e., one with a system, a measuring apparatus and an observer, is only a special case of setting in which the CH formalism allows for probabilities to be computed. Furthermore, it is argued that the probabilities assigned through the CH formalism coincide with the ones one would obtain using the standard interpretation. Consequently, the conclusion is that the standard interpretation is nothing but a special or limiting case of CH; (for a detailed assessment of Griffiths' treatment of this point see \cite{Oko.Sud:13}). 
\subsection{Human language and reality relativism}
In \cite{Har:07} it is argued that most (if not all) of the causes for the uneasiness and discomfort  that underlie the reluctance for accepting CH can be traced to shortcomings in our everyday language. Therefore, the proposal is to explore a possible source of tension between domains in which human language evolved, i.e., quasiclassical realms, and those to which it can be applied, like quantum physics for example. The conclusion reached is that such tension results in the fact that human languages contain \emph{excess baggage} (see also \cite{Har:91a}) that, in order to be useful for physics, must be discarded. As prime examples of excess baggage, the use of the verb `to happen' in special relativity (SR) and in quantum mechanics and of the word `reality' in quantum mechanics are considered.

Let first consider briefly what \cite{Har:07} has to say about the colloquial use of `to happen' in SR. On the one hand, it observes that human language assumes an absolute division of the world into past, present and future. Therefore, it allows constructions of the form: `$A$ happened before $B$' or `$C$ happened at the same time as $D$'. On the other hand, it points out that, according to SR, such absolute division does not exist since the partition of spacetime into past, present and future, depends on the observer. Therefore, absolute statements about the temporal order of events cannot be formulated. To resolve the conflict, the text proposes either to drop all constructions involving `to happen' in a special relativistic contexts, or to use it but with qualifications, as in `A happened before B, in such and such frame of reference'.

As for the use of `to happen' in CH,  it is noted that questions, answers, predictions or retrodictions need the specification of a realm in order to be meaningful. It says, for example, ``If someone asks you `What happened yesterday?’ you should strictly speaking respond with the question `In what realm’.'' However, it recognizes that the colloquial use of `to happen' assumes that only one realm exists so its use must be reformed. Similarly for the use of `reality' since human language assumes that there is only one, but different realms have different notions of `reality'. Therefore, when using the words `real' or `reality' it is necessary to specify which realm is being considered.

In order to fully understand the implications of the proposal, it is necessary to distinguish in it two central claims. The first one consists in holding what could be called a \emph{reality relativism}, i.e., the ontological claim that the notion of reality, or what is real, is meaningful only relative to a realm. This is of course a strong assertion. The second claim maintains that our difficulty for accepting the first one arises from deficiencies in human language. 

DHI introduces some intriguing ideas that might work towards a solution of some of the traditional interpretational problems of standard quantum mechanics. Nevertheless, it suffers of what we take to be a number of severe conceptual problems. We will consider these in section \ref{RO} but before doing so we will review some known objections against CH, along with the standard replies.

\section{``Traditional'' Objections}
\label{MO}
In this section we will briefly review some of the main objections that have been raised throughout the years against the CH formulation of quantum mechanics, along with standard responses. In particular, we will mention four important criticisms of CH that some people have considered devastating. 

The first objection we will mention is related to something we have already discussed: the fact that the theory allows for contrary inferences (see \cite{Ken:97}). As we saw in the previous section, by fixing present data and choosing incompatible past realms, the formalism allows to retrodict contradictory propositions. In fact, as shown there, one can retrodict contradictory facts each with probability one. As we noted above, in order to handle the situation, DHI includes in the formalism a rule forbidding the simultaneous use of incompatible realms in order to make inferences. The objection then consists in claiming that the addition of such a rule constitutes an \emph{ad hoc} solution, void of physical motivations. CH proponents, on the other hand, respond that not combining probabilities from different consistent sets is not an \textit{ad hoc} rule; that its only an application of ordinary probability theory because different realms define different sample spaces (see \cite{Gri.Har:98}).
 
The second objection we will consider is the fact that the theory appears to lack predictive power (see \cite{Dow.Ken:96}). The point is that, in the same way as different past realms can tell different stories of what happened, different future realms can tell different stories of what will happen. 
Therefore, predictions (even those probabilistic in nature) can only be made conditional on a choice of realm. This, together with the fact that the formalism treats all realms on a par, i.e., it offers no procedure of singling out any particular one, seems to imply that there is no way of extracting usable information from the formalism. Advocates of CH argue that once one knows which experiment is being performed, one can fix the realm accordingly. Thus, they argue,  the theory makes predictions for all experiments or observations suitably specified, which is enough for doing physics. Yet, as we will argue below,  taking this view would bring us back to square one because the issue would again be to determine ``under what conditions does the theory specify that a certain experiment is being performed.'' In other words, we would need to solve the measurement problem of quantum mechanics (see \cite{Oko.Sud:13} for a thorough elaboration of this argument).

The third criticism has to do with the fact that, generically, quasiclassical histories cease to be quasiclassical very abruptly (see \cite{Dow.Ken:96}). That is, almost all histories that are quasiclassical up to a point in time, stop being quasiclassical in the future. Therefore, the theory is unable to explain the observed persistence of quasiclassicality. A possible response to this objection holds that quasiclassical realms can be extended into the future with quasiclassical variables. Of course, the response continues, it could happen that in the far future the resulting sets fail to decohere; however, this would not happen abruptly.

The last objection we will mention is related to the fact that the CH formalism delivers approximate probabilities (see \cite{Bar:99}). The problem is that the CH probabilities are approximate but in a very atypical manner. A common way of introducing approximate probabilities into a theory would be through a mechanism which generates results very close to some unknown, but actual probabilities. That would not be that troublesome as long as the formalism guarantees that discrepancies remain small. However, CH probabilities are approximate in a different, much more problematic, fashion because its approximate character implies that they fail, {\it as defined}, to obey the axioms of probability. It is unclear, then, that the numbers provided by the theory can actually be interpreted as genuine probabilities. The typical response maintains that there is always uncertainty in how accurately probabilistic predictions can be checked. Therefore, so as long as the inexactness in the predictions of the formalism, due to the lack of complete decoherence, is maintained below this threshold, no complications will surface. Another option is to only consider formulations of CH where decoherence is satisfied exactly (see, e.g., \cite{Har:08,Gel.Har:11}).

Before moving on, we would like to close this section with a quote that nicely encapsulates the sentiment of critics of the CH formulation, and in particular of the position of DHI:
\begin{quote}
 ``[they] seem - despite much critical probing - unclear on, or uncommitted to taking a stance on, precisely what, if anything, in the theory corresponds to objective external reality'' \cite{Ken:10}.
 \end{quote}
Proponents of CH, on the other hand, believe that all of these issues can be satisfactorily addressed within the CH formulation. We will not attempt here to evaluate either the above mentioned objections or responses (besides what we said about measurements), a lot has been said about them in the literature already. At any rate, in the next section we will discuss what we take to be four novel objections against DHI.
\section{Further Objections}
\label{RO}
In this section we will present four objections against CH that, we believe, turn the formalism, at least in the form advanced in DHI, essentially unsustainable. The first of these objections questions the coherence of the proposal of a realm-dependent reality and concludes, among other things, that by not providing a mechanism for realm selection, the scheme loses its cohesion. Next, we examine the idea that the initial state of the universe, and its dynamics, should be provided by an external theory and test the consistency of the proposal. After that, we challenge the claim that standard quantum mechanics is contained in CH, and we close by dissecting the evolutionary explanation for the existence of IGUSes and their relation to quasiclassical realms. 

\subsection{Realm-dependent reality}

As we saw in the previous section, DHI quite explicitly proposes a reality relativism. The formalism is asking us to consider the idea that reality, or what is real, is relative to a realm. Below we will question the consistency of such a proposal. Before doing so, thought, we will explore if, as \cite{Har:07} asserts, our problem for accepting it emerges from the tension between colloquial language and physics language.

First of all, the history of science is full of examples of concepts that at some point are thought of being absolute but that turn out to be relative. A great example of this, specifically mentioned in \cite{Har:07}, is the fact that in SR the order of events in time is not absolute, as believed within Newtonian mechanics, but relative on the frame of reference. Then, according to SR, it could be the case that for some observer $A$ happens before than $B$, for another $A$ and $B$ are simultaneous and for a third one $A$ happens after $B$; all these even though in Newtonian mechanics temporal order is absolute. Therefore, if true, the reality-relativism proposal would surely not be the first time science discovers something to be relative. Furthermore, human language seems perfectly capable of dealing with relative concepts; we do it all the time with notions like big, far, cold, etc. Therefore,  it is unclear why the uneasiness with the idea of a realm-dependent reality could have anything to do with shortcomings or excess baggage in human language; (in fact, as we will see below, this is not a language problem but an ontologic one).

Another point we would like to stress is that the proposed analogy between SR and CH is not a valid one. The reason is the following: in SR, simultaneity is observer dependent. Nevertheless, the perceptions of all possible observers about the relative order of events (and about anything else) can be codified consistently in a single  mathematically coherent structure: a 4-dimensional manifold, specifically $\mathbb{R}^4$, endowed with a pseudo-Riemannian flat metric, with curves representing the world lines of particles and observers. From such a structure, the point of view associated with each observer can be fully recovered. In this sense, the theoretical framework involves a deeper observer-independent reality underlying the  elements required to deal with the perceptions  of every  observer. This of course ensures the self-consistency of the theory. The trouble with the CH approach is that, in contrast with SR, it offers us no unified and self-consistent model of the world, but only a concatenation of different pictures and a rule that instruct us ``not to use two of them simultaneously.'' Instead, a proposal for a fundamental theory describing our physical world  should be based on a scheme that presents a unified characterization of nature and that it be such that the rules about the use of different sectors of the theory are seen to emerge directly from that picture.

What about the proposal itself for a realm-dependent reality? As we said before, it is, no doubt, a controversial claim. However, before considering the idea seriously, it is necessary to check if it is internally consistent and in accord with the experiences we want to understand within the frame provided by the theory. One of the main problems in this respect is that it is not at all clear how users of the theory, i.e., IGUSes,\footnote{Remember that information gathering and utilizing systems (IGUSes) play a fundamental role in DHI (see section \ref{QI} and references therein). } are supposed to fix or select a specific realm among all the possible options offered by the formalism. Of course, this is an essential step to make sense of the theory. However, the formalism does not explicitly state any clear mechanism for doing so. Furthermore, reading the sources does not help much in clarifying the situation:
\begin{quote}
 ``...we could adopt a subjective point of view... and say that the IGUS ``chooses'' its coarse graining of histories and, therefore, ``chooses'' a particular quasiclassical domain... It would be better, however, to say that the IGUS evolves to exploit a particular quasiclassical domain or set of such domains.'' \cite{Gel.Har:90}.
 \end{quote}
It is not clear, then, whether an IGUS (or a class of IGUSes) chooses a realm, or weather it is the realm that limits or constrains the possibility of existence, and characteristics, of IGUSes dwelling within it. In any case, there are just two basic options: either IGUSes can or cannot choose realms. The problem is that neither option seems to takes us to satisfactory conclusions. If selecting a specific realm is beyond our capacities as IGUSes, then talk of multiple realms seems extravagant and serves no real purpose in the theory (other realms being empirically inaccessible). Furthermore, if it is not the IGUS that does the choosing, what is the entity or circumstance that does it and how does it do it? On the other hand, if an IGUS can choose a realm, proponents of the formalism owe us an explanation of how this could be so,\footnote{Recall that we already indicated that one cannot argue that the experimental set up is what determines the choice because then the issue would again be to specify under what conditions does the theory indicate that a set up counts as an experiment (see \cite{Oko.Sud:13}).} especially after noticing that it involves fixing projections everywhere in the universe, and at all times, and, moreover considering that the corresponding projections might radically affect our current experience, or even alter the fact that we exist in the present, (see \cite{Sud:11}). The problem then is not only that a mechanism for selecting a realm is missing, but that the formalism seems to lack the resources for providing it.

A related complication is the following. Proponents of CH maintain that a realm must be chosen according to the questions one is trying to answer and the predictions one is interested in obtaining. However, the issue that concerns us here goes in a different direction. We are not interested in a recipe for applying the formalism in order to come up with predictions for experiments in which we can take for granted a myriad of things - like a distinct system, a measuring apparatus, well defined observables, observers, etc. We are rather interested in evaluating the formalism as a theory of all these things together, which of course is the central motivation for taking it seriously in the first place; in fact  the  main motivation for  proposing something like   the CH  formulation is  precisely the fact that the Copenhagen   interpretation is only tenable (if at all)  when  one can take  that myriad of things as given. We believe then that there are two different levels of discourse that get entangled: one is about how IGUSes use the theory to make predictions and the other is about what the theory tells us with respect to the nature and functioning of the world as a whole.\footnote{An instrumentalist with respect to quantum mechanics could object that this distinction does not make sense but proponents of CH must recognize it because it is central to the whole motivation for the approach.}  We also believe that it is of extreme importance, for an adequate assessment of the CH approach, to always be clear about the distinction between these two levels of discourse. Actually, regarding this last issue, one has to wonder how, if our world is in fact accurately described in terms of the CH formalism, could IGUSes such as ourselves ever be able to come up with quantum theory in general, and with the CH approach to it in particular (see \cite{Bar:96} for an interesting exploration of this question).

\subsection{Initial conditions}

The second objection we would like to mention is related to the idea that the CH formalism takes as inputs an initial state $\ket{\psi}$ and a Hamiltonian $\hat{H}$. These objects, which are necessary for making predictions and retrodictions (see equations \ref{pre} and \ref{ret}), are supposed to be fixed and absolute, i.e., realm-independent. However, if the idea of a realm-dependent reality is taken seriously, it is far from clear how one could have access to this absolute elements ($\ket{\psi}$ and $\hat{H}$). In other words, how are we supposed to choose initial conditions for a theory that holds that the present does not uniquely determine the past? As a way out of this state of affairs, DHI proponents insist that we need to construct a separate and external theory for choosing the initial conditions (see \cite{Gel.Har:90,Har:91,Har:96,Har:97,Har:98,Har:03,Har:05}). However, if the past really is relative, nothing at all that we observe can ever count as evidence for such an external theory. That is, we cannot, even in principle, test such ideas about the initial state.

One could argue that since CH contains standard quantum theory as a special case (something we question below), then if one can determine the initial state and the Hamiltonian in standard quantum theory one should be able to do so in CH. However, the key element of our objection is the “reality relativism” proposed by DHI. Since standard quantum mechanics does not share that characteristic with DHI, the counterargument loses its force. 

A more thoughtful objection%
\footnote{We thank an anonymous referee for raising this particular objection and for urging us to develop further our argument.}
claims that since the initial state and the Hamiltonian can be determined in the standard quantum case they can also be determined in a \emph{single} realm in CH (the one one is using while doing the measurements and calculations). But since the state and Hamiltonian are ``realm-independent'' then determining them in any realm fixes them in all realms, so we are done. The problem with this argument is that instead of being considered a \emph{proof} for the possibility of finding ``realm-independent'' initial state and Hamiltonian, it should be recognized as a \emph{demonstration for the impossibility} of doing so. That is because, as a matter of fact within the CH framework, \emph{the past is relative to the realm} (see \cite{Har:98a}). And so, whatever one takes to be the initial state in a given realm, will not be so in a different one.


\subsection{Measurements}

We turn next to two related issues, both having to do with the treatment of the concept of measurements in CH. The first is whether, as claimed, the formalism contains standard quantum theory as a special case; the second is related to the fact that DHI fails to establish a clear link between actual (physical) measurements and the projection operators of the (mathematical) formalism. Let us expand on this.

As we saw in section \ref{GHI}, DHI sustains that the CH formalism incorporates Copenhagen quantum theory as a limiting scenario. If this is true, it of course implies that CH is consistent with experiments (to the extent that standard quantum mechanics is). However, the situation is a bit more complicated than what is suggested there. The first problem is that the
scheme offers no way to decide, even after (somehow) fixing a realm, what is the status of the different histories within it. Once again we see two available options:
\begin{enumerate}
  \item Only one of the histories within the chosen realm is actual, in which case the formalism is descriptively incomplete since, as it lacks a reduction postulate to tie outcomes to histories, it does not offer a mechanism to explain the preference for the chosen history among the alternatives. 
In other words, it does not ascribe to the actual story any ontological status, and thus no special role whatsoever. 
  \item All the histories within the chosen realm are actual, in which case there are two problems. On the one hand, it is not possible to interpret as probabilities the numbers generated by the framework since all options are realized. On the other, it sharply conflicts with our everyday experience of obtaining determined outcomes when we perform measurements, (these of course are the standard objections against many-world scenarios).\footnote{Unlike DHI proponents, Griffiths unambiguously states that the adopts the first option above, \cite{Gri:03}.}
\end{enumerate}

Another important omission in DHI with respect to measurements is that, as we briefly mentioned before, the proposal does not make explicit what is the relation between measurements performed by IGUSes and the projection operators of the formalism. That is, there is no specification of how are we to connect the mathematical formalism provided with experimental practices, (Born's rule plays this role in the standard interpretation but of course the idea of any alternative to the Copenhagen interpretation is to improve upon it). The only rule that the formalism provides is the following: realms are to be chosen according to the questions one is trying to answer. The issue we would like to examine then is how one is supposed to apply this dictum in practice. That is, given a standard measurement situation, which is exactly the realm one must use? 

An initial (and partial) response to the question raised above could be that the realm must contain, as a minimum, a projection corresponding to the measurement to be realized. This, however, is deeply problematic because, as we said before, the CH formalism does not specify under what conditions one is allowed to conclude that a measurement is taking place (in other words, the old measurement problem that motivated  going beyond the Copenhageun interpretation crops up once again).

The situation is even worse for the rest of the projections that comprise the realm (remember that realms contain sets of projections at various times). Clearly, these cannot be associated with measurements performed by IGUSes because for a measurement situation to arise, specific projections must had happened early on in the history of the universe, before any IGUS was around to perform measurements. The remaining option is to disassociate the projections of the formalism with measurement but this solution is also unacceptable because the formalism lacks resources to do so. That is, it does not posses any other element that could do the job of relating experiments to the formalism.

\subsection{IGUSes and quasiclassicality}

The last critique we will offer questions the viability of the proposed evolutionary explanation for the existence of IGUSes and their relation to quasiclassical realms. As we saw in section \ref{GHI} the idea is that it is evolutionarily advantageous to be able to make predictions and so IGUSes are selected for because they are good at it. Furthermore, they evolve in quasiclassical realms because these are the environments that present enough regularities so that predictions can be generated. There are, however, serious problems with this reasoning. 

Lets start by asserting that evolutionary theory minimally includes the following: \emph{impact of the environment, { and the fitness  relative to said  environment of the organisms in question, on  their } reproductive success}. Therefore, its essential elements include: a varied initial population, an external environment, heredity and selection. However, none of these elements seem to be present in the CH context. In particular, there is no external given framework for IGUSes to evolve since, according to the proposed explanation, the environment (i.e., the realm) is an essential part of what is supposed to be adaptively selected (the result purportedly being a quasiclassical realm). In other words, one cannot argue that evolution takes place, according to the standard paradigm of natural selection, unless one can argue that things do occur: that a failure to obtain resources results in death, or that organisms that are unfit do not reproduce, etc. Those rules presuppose a quasiclassical realm, and it is thus clear that they cannot be used to argue that they play a role in selecting one such realm over something else.

For comparison, we note that the \emph{anthropic principle} can be stated as follows: features of the world are what they are because, otherwise, we wouldn't be here to remark on it. This, we believe, sounds a lot closer to what is being proposed by DHI since, in effect, what is being argued is that we experience a quasiclassical realm because it is the only one that allows for IGUSes like ourselves to exist. The purpose of this observation is not to question the usefulness or validity of the anthropic principle, this is not the place for doing so. The objective is to demonstrate that what is presented in DHI as an evolutionary argument is, at best, an invocation of the anthropic principle.
\section{Conclusions}
\label{Conc}
There is no doubt that the attempt to develop a generalization of standard quantum theory, applicable to closed systems, is an important enterprise; even an essential one when viewed as a foundational step in the construction of quantum theories of cosmology. It is also a very interesting and worthwhile project to explore whether purely unitary quantum theory can be cast into a scientifically adequate theory. That is, if a version of quantum theory where temporal evolution is implemented purely in terms of Schr\"odinger's equation, with no mention of a  reduction postulate (or something else playing a similar role), can be made compatible with our experience of definite measurement results and a stable characterization of the world and the laws of nature. On the other hand, it is very likely that the problems encountered while trying to apply quantum theory to the universe as a whole arise not from our inability to interpret quantum theory, as advanced by CH proponents, but from the (bold) assumption that quantum theory is universally valid.

In any case, before loosing hope on a theory as successful as quantum theory, we think it is wise to explore how far it can be extended. The formulation of CH proposed in DHI surely is a brave attempt in this respect, unfortunately, at least in its present form, it cannot be considered as truly satisfactory. 

\section*{Acknowledgements}

We would like to acknowledge partial financial support from DGAPA-UNAM projects IN107412 (DS), IA400312 (EO), and CONACyT project 101712 (DS).


\end{document}